\documentclass[journal]{article}
\usepackage{amsmath}
\usepackage{amssymb}
\usepackage{amsfonts}
\usepackage{fullpage}
\usepackage{graphicx}
\usepackage{float}
\usepackage[center]{caption}
\usepackage{subcaption}
\usepackage{multirow}
\usepackage{textcomp}
\usepackage{wasysym}

\begin{document}

\title{Radiological images and machine learning: trends, perspectives, and prospects}

\author{Zhenwei Zhang, Ervin Sejdi\'{c}\thanks{Zhenwei Zhang and Ervin Sejdi\'{c} are with the Department of Electrical and Computer Engineering, Swanson School of Engineering, University of Pittsburgh, Pittsburgh, PA, 15261, USA. E-mail: esejdic@ieee.org. Ervin Sejdi\'{c} is the corresponding author.}}

\date{}
\maketitle

\begin{abstract}
The application of machine learning to radiological images is an increasingly active research area that is expected to grow in the next five to ten years. Recent advances in machine learning have the potential to recognize and classify complex patterns from different radiological imaging modalities such as x-rays, computed tomography, magnetic resonance imaging and positron emission tomography imaging. In many applications, machine learning based systems have shown comparable performance to human decision-making. The applications of machine learning are the key ingredients of future clinical decision making and monitoring systems. This review covers the fundamental concepts behind various machine learning techniques and their applications in several radiological imaging areas, such as medical image segmentation, brain function studies and neurological disease diagnosis, as well as computer-aided systems, image registration, and content-based image retrieval systems. Synchronistically, we will briefly discuss current challenges and future directions regarding the application of machine learning in radiological imaging. By giving insight on how take advantage of machine learning powered applications, we expect that clinicians can prevent and diagnose diseases more accurately and efficiently.


\end{abstract}

\section{Introduction \label{sec: introduction}}

\begin{figure}[!h]
 	\centering
 	\begin{subfigure}{0.3\textwidth}
 		\centering
 		\includegraphics[height=3cm]{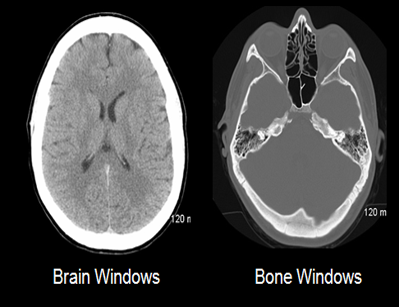}\caption{\label{fig:a}}	
 	\end{subfigure}
 	\begin{subfigure}{0.3\textwidth}
 		\centering
 		\includegraphics[height=3cm]{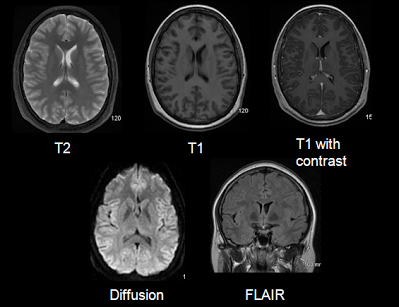}\caption{\label{fig:b}}
 	\end{subfigure}
  	\begin{subfigure}{0.3\textwidth}
 	\centering
 	\includegraphics[height=3cm]{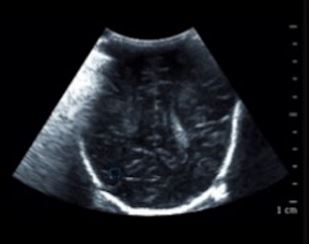}\caption{\label{fig:c}}
	\end{subfigure}
 	\caption{An example of CT (a), MRI (b) and ultrasound (c) images displaying brain structures. Soft tissue has a better resolution in MRI images. Each types of MRI sequence displays a different brightness for the same structures \cite{CTimage}. Ultrasound is more convenient than CT and MRI, however it is unable to capture information well, as ultrasound waves do not transmit well through bone \protect\cite{bailey2017contrast}. }
 	\label{fig: result_all}
 \end{figure}

\begin{figure}
	\centering
	\includegraphics[width=2in]{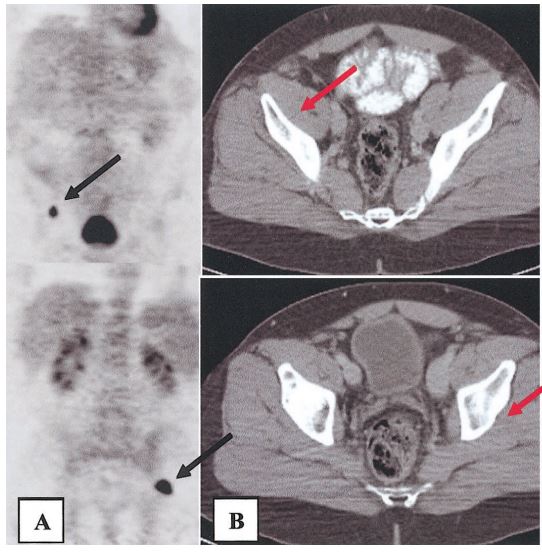}
	\caption{(A) axial views of a CT scan, (B) coronal PET. CT images show better resolution than PET images. However, each type of image can provide useful information for diseases. In this case, coronal PET images shows multiple foci of intense FDG uptake in the pelvic area while CT images do not demonstrate any abnormalities \protect\cite{townsend2003pet}}
	\label{fig:PET}
\end{figure}

Radiology is a branch of medicine that uses imaging techniques to detect, diagnose and treat diseases \cite{novelline2004squire,chen2010basic,herring2015learning}. Diagnostic radiology helps radiologists image internal body structures to diagnose the cause of symptoms, screen for illnesses and detect the body's response to treatments. 
The most common radiology modalities include: plain X-ray, computed tomography (CT), magnetic resonance imaging (MRI), positron emission tomography (PET), and ultrasound imaging. 
Fig. \ref{fig: result_all} shows these internal body structures viewed via these different imaging techniques, and Fig. \ref{fig:PET} illustrates an example of CT and PET images. In MRI images, the white areas represent subcutaneous fat, while in the CT images, the white areas represent the skull. However, the main disadvantage for all x-ray and gamma ray imaging modalities is the risk of radiation exposure for patients \cite{ Swensen2005,Iyer2010,Pearce2012,Smith-Bindman2009,Frush2003}.
Ultrasound imaging is convenient because it does not expose patients or radiologists to radiation, but it has poor penetration through bone or air, which makes images difficult to interpret \cite{Huang2004,Shan2015}.
 MRI and CT images can capture anatomical changes in tissues, while PET images detects biochemical and physiological changes, which often occur before anatomical changes \cite{Wang2016a}. Disadvantageously, patients with ferromagnetic orthopedic implants, materials, and devices cannot undergo MRI procedures. MRIs also have relatively long scanning times which imposes limitations for patients in need of urgent care \cite{Sundaram1988, Dong2015}.  The broader use of radiological image analysis increases the workload for radiologists, and therefore the development of intelligent computer-aided systems for automated image analysis that can achieve faster and more accurate results for large volumes of imaging data is essential.

This paper provides an overview of machine learning techniques used in radiological image analysis. We begin with a brief overview of current imaging technologies. In section \ref{sec: Machine Learning in Radiology}, we review general concepts of machine learning and detail methods most commonly used in recent years. In section \ref{sec: Application of Machine Learning in Radiology Images}, we provide an overview of the most current studies dealing machine learning and radiological images. This review paper mainly focuses on the most recent contributions to different machine learning techniques (i.e., after 2014), and the reader should refer to previous review papers for older contributions related to machine learning and biomedical imaging \cite{Ibrahim2016,Wang2013,Bauer2013,Garcia-Lorenzo2013,Kourou2015}, or contributions that focus solely on a single machine learning approach (e.g., deep learning \cite{litjens2017survey, shen2017deep}). Lastly, we have summarized these contributions by outlining current technological limitations and potential future areas of research in this field.  

Contributions cited in this review were collected using various research databases such as GoogleScholar, PubMed (MEDLINE), IEEE Xplore and SpringerLink.  All contributions collected were published between the middle of 2014 and the middle of 2017. We used variations of the keywords including but not limited to combinations of machine learning techniques (SVM, random forest, regression, neural networks, deep learning), applications (segmentation, computer assisted system, brain studies) and imaging modalities (MRI, x-rays, ultrasound, CT). While deep learning techniques have been prevalent in the past five years, our search not only included these hot topics but also included traditional methods. In this review, the preference was given to papers that presented real data rather than theoretical frameworks. Similarly, we did not include papers that repeated past experiments unless the data collection or data analysis procedures were different.

\section{Machine Learning in Radiology \label{sec: Machine Learning in Radiology}}

In recent years, machine learning algorithms have become useful tools for the analysis of medical images in many radiology applications \cite{Wang2013,erickson2017machine}. For example, machine learning algorithms can extract the useful information found within the details of medical images  \cite{Salvatore2014}. Thus, computer-aided systems based on machine learning help radiologists to make informed decisions while interpreting these images \cite{Wang2013}.

\subsection{Types of Learning}
Depending on the utilization of labels in training data, there are three categories of machine learning algorithms: supervised learning, unsupervised learning, and semi-supervised learning.
Supervised-learning is the most common form in machine learning, and researchers widely use supervised-learning for classification and regression \cite{Rusk2015}. Data is usually collected and labeled in categories, as the purpose of supervised learning is to find an appropriate input-output function from training data, which generalizes well against the testing data. We can compute an objective function to measure the error between the desired pattern and the output score. In general, many scientific contributions focus on finding a suitable objective function with adjustable parameters. 
Contrariwise, in cases where labeled data sets are relatively rare or difficult to acquire, unsupervised learning can derive deductions from data without corresponding label information; the purpose of unsupervised learning is to discover the hidden structure or distribution of data \cite{Dayan2009}. Unsupervised learning approaches include clustering and blind signal separation techniques such as principal component analysis and independent component analysis.
Lastly, semi-supervised learning lie between supervised learning and unsupervised-learning \cite{Mitchell1998, Zhu2011}. During the training phase, semi-supervised learning begins with a small set of labeled data and augments the training data size by gradually labeling unlabeled data.

\subsection{Feature Selection}
Feature extraction and representation is a crucial step in medical image processing. With the development of modern medical techniques, higher resolution and more features have become obtainable to feed the classifiers; however, this is an obstacle for machine learning techniques in achieving an optimal solution using high dimensional features. Significant interest exists in extracting and identifying reliable features from radiological images to improve classification performance\cite{Chao2013, Larroza2015}. Several methods exist for extraction of features from medical images including region-based, shape-based , texture-based, and bag-of-words features \cite{Tsai2007, Islam2008, Yang2008, Tian2013, Xie2016, Rastghalam2016, Meng2016}. The performance of most image retrieval systems is dependent on the use of these features. Table \ref{segmentation table} summaries image features used in radiological image analysis. Color features are one of the essential features of images, including RGB, histograms \cite{Yue2011}, color moments \cite{Pass1996} and color coherence vectors. Groups of pixels can calculate texture features, which can help characterize a wide range of images. The Gabor filter is the most common method for texture extraction \cite{Tian2013}. Scale invariant feature transform and speed up robust features algorithm are two popular methods for scale and rotation invariant feature detector and descriptor in computer vision\cite{Juan2009a}. Different types of images have significant contrast variation. Thus visual features such as color, shape and texture are not enough to easily classify images. Thus high-level features are useful to overcome the intensity variations in different types of images and extract the appropriate information from said images. The process to select ideal features that can reflect the most useful contents of images remains a challenging problem in machine learning.  

\begin{table*}[t]
	\centering
	\caption{A summary of image features used in ML systems}
	\label{segmentation table}
	\begin{tabular}{lll}
		Features &                                                       & Examples                                            \\ \hline
		Color    & Invariant from different size and direction           & Histogram \cite{Srinivas2015, R.R.2015, Chen2016}   \\
		Shape    & Binary representation of images                       & Sphericity \cite{Chen2016,Dhara2016}                \\
		Texture  & Description of image structure, randomness, linearity, & Haralick's features \cite{Suresh2015,Dhara2016}     \\
		& roughness, granulation, and homogeneity                 & Gabor features \cite{Keserwani2016,Zhu2016,Lee2016} \\
		&                                                       & Co-occurrence \cite{Murala2013}                              \\
		&                                                       & Curvelet-based \cite{Dhahbi2015,Sethi2015}          \\
		&                                                       & Wavelet-based \cite{Pereira2014,MaderoOrozco2015}   \\
		Local    & Description of local image information using region,  & Local binary pattern \cite{Chen2016}                \\
		& object of interest, corners, or edges                  & Scale invariant feature transform \cite{Arias2016,Lee2016a,Alkhawlani2015}                      \\
		&                                                       & Speed up robust features \cite{Velmurugan2011,Alkhawlani2015}                         \\
		Other    & Other methods to extract image features               & CNN \cite{Srinivas2016a}                           
	\end{tabular}
\end{table*}

\subsection{Overview of Machine Learning Methods \label{sec: Overview of Machine Learning Methods}}
Machine learning has been developing rapidly in recent years, and it is impossible to cover all recently-developed techniques in one section. In this section, we will review the most commonly used machine learning methods in radiology, such as linear models, the support vector machine, decision tree learning, the ensemble classifier, as well as neural networks and deep learning. This section provides a general description of machine learning techniques and will help understanding their applications in the field of radiology, as described in subsequent sections.

\begin{figure}[!h]
	\centering
	\begin{subfigure}{0.2\textwidth}
		\centering
		\includegraphics[width=0.99\linewidth]{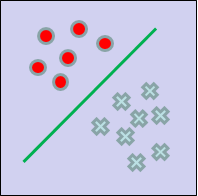}\caption{\label{fig:5a}}
	\end{subfigure}	
	\begin{subfigure}{0.2\textwidth}
		\centering
		\includegraphics[width=0.99\linewidth]{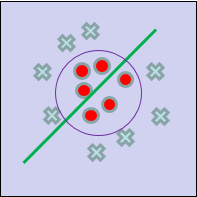}\caption{\label{fig:5b}}
	\end{subfigure}
	
	\caption{Basic idea of linear classification and non-linear classification, (a) linear case (b) non linear case. The linear model uses linear functions to separate the data yet is not suitable for non-linear cases. SVM is one way to separate non-linear models using different kernel functions.}
	\label{fig：variance}
\end{figure}

\subsubsection{Linear Models for Regression and Classification}
Regression predicts the value from the given input features, whereas classification assigns input $x$ to one of the predefined classes.\cite{Bishop2006}. The simplest linear models establish a linear relationship among input variables. Commonly used linear models include linear regression, Fisher's linear discriminant (LDA), and logistic regression. 
Given $x_i,i=1,2,3...,N$, the input feature vector, the output $y(x,\omega) = \omega_0+\sum_{i=1}^{N}\omega_Nx_N$. Logistic regression is the most basic classifier, it predicts the probability that an input $x$ belongs to a class (class 1), versus the probability that it belongs to another class (class 0). The basic idea of logistic regression is that we learn the logistic function of the form:
\begin{equation*}
P(y=1|x)=\frac{1}{1+exp(-\omega^Tx)}
\end{equation*}
where $x$ is the input vector and $\omega$ is a weight vector for input. The logistic function is a continuous function which can turn any input from negative infinity to positive infinity into an output that is always between zero and one \cite{Deng1998}. Fig. \ref{fig：variance} illustrates linear and non-linear separable cases for a dataset.

\subsubsection{Support Vector Machine}
Support vector machines (SVM) are kernel-based supervised learning techniques widely used for classification and regression \cite{Bishop2006,Suykens1999}. The basic idea of SVM is to find an optimal hyperplane for linear separable patterns. It attempts to maximize the geometric margin on the training set and minimize the training error. Then, a kernel function maps the original data into a new space for non-linearly separable cases, resulting in a two-class classification problem.  $x_i,i=1,2,...,N$ are feature vectors of the training set $X$, and of corresponding class indicator $y_i \in \{-1,+1\}$. The goal of SVM is to construct a classifier in the form of: 
\begin{equation*}
y(x)=\text{sign}[\sum_{i=1}^{N_s}\lambda_i y_iK(x_i,x)+\omega_0]
\end{equation*}   
The function $K(x_i,x)$ is called the kernel function, and their different mathematical properties enable many pattern recognition and regression models.
SVM with a linear kernel equation is computationally faster than SVM with quadratic kernel functions. SVM models using fewer but more significant features are most likely robust and less prone to overfitting \cite{Torheim2014}.

\begin{figure}
	\centering
	\includegraphics[width=3.5in]{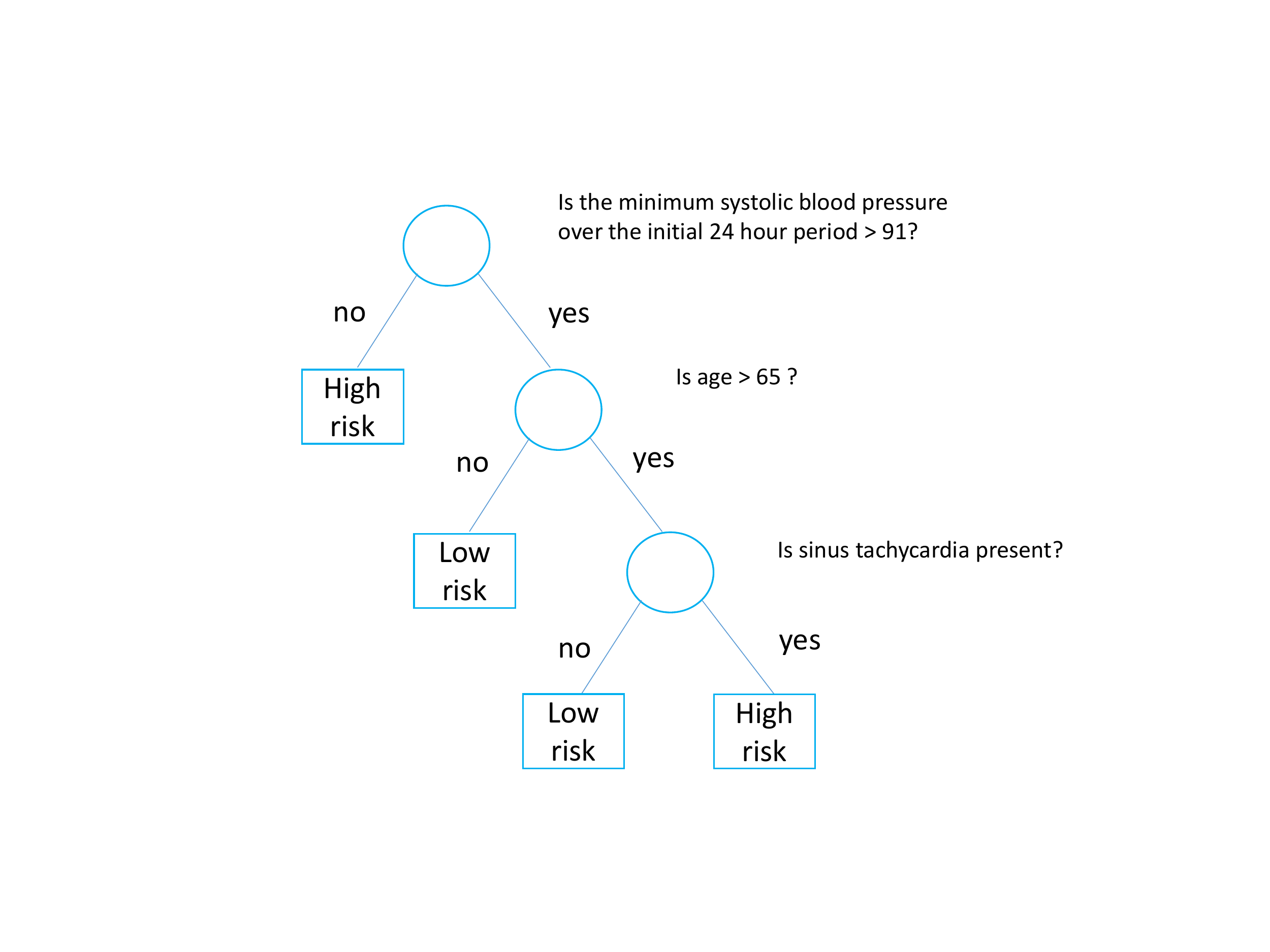}
	\caption{A medical example of decision trees. In this example, patients are classified into two classes: high risk and low risk. The features include blood pressures, age, etc. In this case, the classification tree operates similarly to a clinician's examination process.}
	\label{fig:DT}
\end{figure}

\subsubsection{Decision Tree Learning}
Decision trees are one of the most popular classification approaches in machine learning \cite{Loh2014}. The decision tree consists of  a ``root'', "leaves", and internal nodes \cite{Rokach2010,Azar2013,Speybroeck2012a}. The internal nodes use certain features to split the instance space into two or more subspaces. Each leaf represents one class. The leaf may represent the most appropriate target value or indicate the probability of the target having a specific value. Fig. \ref{fig:DT} is an example of the decision tree model. Decision trees are capable of handling datasets that may have missing values and errors, however, this method may overfit training data and add unnecessary features. In radiological image analysis, decision trees are usually ensembled to form random forests for prediction and classification.

\subsubsection{Ensemble Learning} 
\begin{figure}
	\centering
	\includegraphics[width=2in]{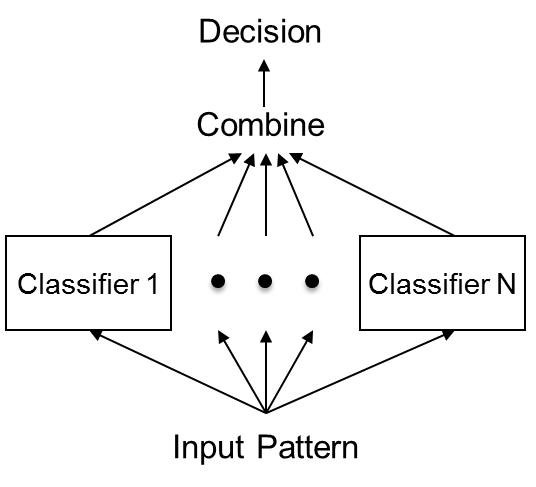}
	\caption{The concept of ensemble learning: an ensemble classifier is made up of several sub-classifiers, the final output is combined with outputs from these sub classifiers and their weights.}
	\label{fig:ensemble}
\end{figure}

Ensemble learning combines multiple classifiers and applies voting algorithms to achieve a final classification. Popular ensemble approaches include boosting and bagging \cite{Bauer1999}. Fig. \ref{fig:ensemble} shows the basic idea of ensemble learning. In boosting, extra weight is assigned to incorrectly predicted points, and a set of weak classifiers are applied to deal with data in the training phase; the outputs of weak classifiers and the weighted inputs help calculate the final prediction. In bagging, the sub-classifier is independently constructed using a bootstrap sample of the data set and a majority voting method is applied for the final prediction \cite{Liaw2002}. Random forests are an ensemble learning method that consists of a multitude of decision trees. In standard tree construction, the node is split using the best split among all features. In a random forest, a random subset of features split each node. The random forest is one of the most powerful machine learning predictors used in detection, classification, and segmentation \cite{TriHuynh2015}, particularly for brain \cite{Zikic2012, Geremia2011} and heart \cite{Sammouda2012, Lempitsky2009} images.

\subsubsection{Neural Networks and Deep Learning}

Deep learning techniques have become a hot topic in machine learning due to the availability of sufficient computational power and a high volume of data. These approaches can select specific features directly from the data for classification and detection purposes\cite{Shin2016, Song2015}. Deep learning avoids designing specific features from the data, which is its main advantage in comparison with other machine learning methods. Some outstanding frameworks such as the restricted Boltzmann machine \cite{Salakhutdinov2009}, convolutional neural networks (CNNs) \cite{LeCun2010} and sparse autoencoders have proven useful tools in many applications such as Alzheimer's disease diagnosis \cite{suk2013}, segmentation \cite{Tong2013}, and tissue classification \cite{Cruz-Roa2013}.  CNNs have a large number of parameters, which requires vast volumes of labeled training data. However, this requirement makes the training of CNNs from medical images challenging due to the difficulty of acquiring a database with labeled data \cite{Shiraishi2009}. However, several studies use CNNs to extract features for medical images and achieve good performance in classification \cite{VanGinneken2015, Choi2015}. 

\subsection{Evaluating Machine Learning Techniques }
Physicians may rely on the prediction or classification results of machine learning algorithms. However, performing one round of training and testing on data sets may not yield a meaningful idea of the accuracy of an algorithm. Cross-validation reduces the variance of accuracy scores by ensuring that each data instance is used for both training and testing an equal number of times. The cross-validation method randomly splits data into $k$ subsets and holds out each one while training on the rest.

\begin{figure}
	\centering
	\includegraphics[width=3in]{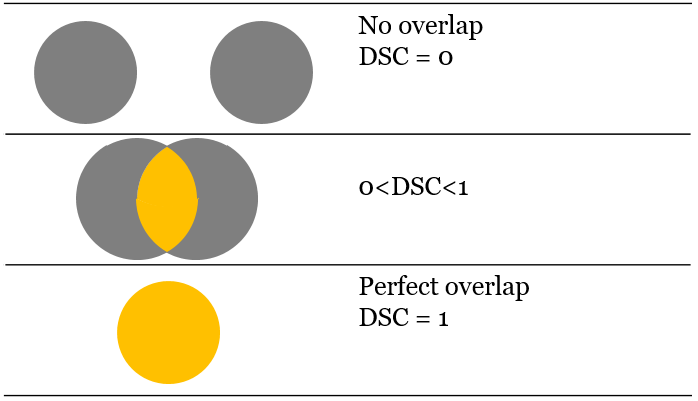}
	\caption{The Dice similarity coefficient represents spatial overlap.}
	\label{fig:dsc}
\end{figure}

The Dice similarity coefficient is used in segmentation, and it measures the spatial overlap between two segmented target regions\cite{Zou2004}. A and B are target regions or volumes, and the Dice similarity coefficient is defined as the ratio of their intersection to the average \cite{Reed2009}:
\begin{equation*}
DSC(\text{A,B}) = \frac{2(\text{A}\cap\text{B})}{\text{A}+\text{B}}
\end{equation*}
 The Dice similarity coefficient has a value of 0 for no overlap and 1 when pa complete agreement is present. Fig. \ref{fig:dsc} illustrates the Dice similarity coefficient with different overlaps.

In clinical practice, subjects with a disease are labeled as positive and  healthy subjects are labeled as negative. True positive (TP), false positive (FP), false negative (FN) and true negative (TN) are defined as follows:
 
TP: a test detects the disease when the disease is present

TN: a test does not detect the disease when the disease is absent

FP: a test detects the disease when the disease is absent

FN: a test does not detect the disease when the disease is present

The goal of a computer-aided diagnosis system is to detect as many true positives as possible and minimize false positives and false negatives. There are several popular metrics used to assess classifier outcomes. Sensitivity shows the ability of a test to correctly detect patients with diseases while specificity is the ability of a test to identify healthy subjects correctly. 
 They can be written as:
\begin{equation*}
\text{sensitivity} = \frac{\text{True Positive}}{\text{True Positive + False Negative}} 
\end{equation*}
\begin{equation*}
\text{specificity} = \frac{\text{True Negative}}{\text{True Negative + False Positive}} 
\end{equation*}

\begin{figure}
	\centering
	\includegraphics[width=3in]{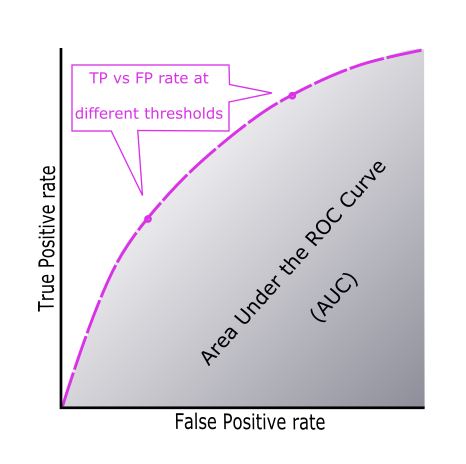}
	\caption{ROC curves consist of the points evaluated from model many times with different classification thresholds. AUC computes the area beneath the ROC curves, which is more efficient to evaluate the models compared to ROC curve.}
	\label{fig:ROC}
\end{figure} 

Other popular methods used to assess models include the area under the receiver operating characteristic (ROC) and the top precision value. ROC curves describe the relationship between sensitivity and specificity. The area under the curve (AUC) measures the entire area under the ROC curve from (0,0) to (1,1) and represents the probability that the model can distinguish between classes. Figure \ref{fig:ROC} illustrates the relation between ROC and AUC. The top precision, the portion of top-ranked relevant images before the top irrelevant database image \cite{Meng2016, Li2014a}, is a popular evaluation metrics in retrieval systems.

\section{Application of Machine Learning in Radiology \label{sec: Application of Machine Learning in Radiology Images}}
\subsection{Segmentation}
Image segmentation is a necessary step in effective disease diagnosis and treatment in radiology imaging research. It helps clinicians to understand structural information and spatial
anatomic relationships, however, it depends on the experience of clinicians and is very time-consuming \cite{Salvatore2014}. Automatic classification methods are essential for improving diagnosis analysis and for the reproducibility of large-scale clinical studies.

\subsubsection{Brain segmentation}
Tree based methods are hot topics currently being investigated in the brain segmentation field. For example, Yoo et al.  segmented multiple sclerosis lesions in multi-3D MR images from unsupervised features\cite{Yoo2014}. Features were extracted from T2-weighted and proton density MR images using a deep belief network, and a random forest was built for the final supervised classification. In order to improve the model performance from noisy training data and robustness against overfitting, Maier et al. proposed an extra tree forest to locate, segment and quantify sub-acute ischemic stroke lesions \cite{Maier2015}. They used voxel-wise local features such as intensity, weighted local mean, local histogram and 2D center distance. However, their method can only deal with the T1-weighted and diffusion-weighted data sequences and high-quality images. Multimodal data from the same patient can provide extra useful information for diagnosis. Therefore, Mitra et al. proposed to use features from multimodal data to segment ischemic lesions, white matter and other secondary lesions. In their study, algorithms combined expectation maximization likelihood estimation and Bayesian-Markov random field to segment the probable lesion areas from FLAIR data then applied random forest on the multimodal data \cite{Mitra2014}.  

\begin{figure}
	\centering
	\includegraphics[width=3in]{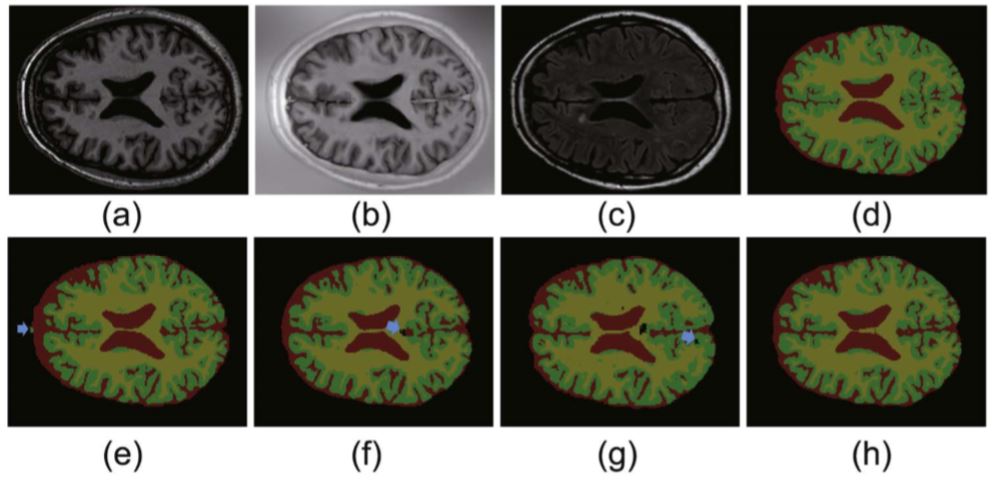}
	\caption{Chen et al. applied their model on different imaging modalities: (a)-(c) denote T1, T1-IR and T2-FLAIR MR images; (d) represents the ground truth label; (e)-(g) illustrates the segmentation results using single image modality respectively; (h) is the result that combines all image modalities.  \protect\cite{chen2017voxresnet}.}
	\label{fig:chen}
\end{figure}

 Neural networks and deep learning techniques are powerful tools in brain segmentation tasks. Si et al. proposed a semi-automatic method to classify the pixels of brain MRI into lesioned and healthy tissues by use of an artificial neural network with gray levels and statistical features as inputs \cite{Si2016}. The segmentation of early-brain tissues is more difficult than that of adult brains due to the lower tissue contrast \cite{Li2011}, while multiple image modalities contain complementary information for insufficient tissue contrast \cite{Wang2014}. Zhang et al. \cite{Zhang2015} showed that fractional anisotropy images are more potent in distinguishing gray matter and white matter, and that T2-weighted images have higher performance in capturing cerebrospinal fluid. Zhang et al. proposed a CNN method combining these multiple modality image data to improve segmentation performance. Similarly, Kleesiek segmented the brain and non-brain tissues by feeding data into a neural network with seven hidden convolutional layers \cite{Kleesiek2016}. Their model can be applied on any single image modality or a combination of several modalities with varying size. Deep learning methods can also automatically segment MRI images of the human brain into many anatomical regions \cite{de2015deep,moeskops2016automatic }. As shown in Fig \ref{fig:chen}, Chen et al. extended ResNet into volumetric brain anatomical segmentation \cite{chen2017voxresnet}. They integrated the low-level image appearance features, implicit shape information and high-level context to further improve the volumetric segmentation performance \cite{chen2017voxresnet}. 

\subsubsection{Other segmentation applications}
Segmentation is also applied to identify and detect other structures \cite{Lindner2013}, such as organs, bones, muscles, and fractures.  Similar to the brain segmentation, tree-based methods are popular as well in other types of segmentation tasks. Lombaert et al. presented kidneys segmentation using the Laplacian Forest \cite{Lombaert2014}. They used intensity within a randomly-shaped cuboid centered around several pixels during their data training. The idea of the Laplacian Forest is to use a guided bagging strategy to produce more related image information for tree models, which have more substantial improvements in model accuracy. Conze proposed a semi-automatic liver tumor segmentation combining a simple linear iterative clustering super pixel algorithm and random forest, which considers the inter-dependencies among voxels \cite{B2015}. The multi-phase cluster-wise features that consider the spatial consistency applied in their approach are more robust for a random forest.
The analysis of the knee also plays vital role in clinical assessment and surgical planning of the disease. The cartilage is typically small, and the segmentation results of Haar-like operators are often unreliable in extracting context features. To overcome these limitations, Liu proposed a novel method using a multi-atlas context forest, which segments bones first and then cartilage \cite{Liu2015}. They trained classifiers using appearance features and context features to align the expert segmentation of the atlases in each iteration.

Medical segmentation research utilizes regression-based models. Chen et al. proposed an automated method to localize and segment intervertebral discs from MRI \cite{Chen2014a}. They used unified regression and classification frameworks to estimate displacements for image points by using the visual features around them and achieved satisfactory results. Ventricle structure segmentation in MRI is an essential task for investigating most cardiac disorders. The primary challenge of this task is the considerable shape variation among different patients \cite{Sedai2015}. et al.  proposed a segmentation method using cascade shape regression for the right ventricle in cardiac MRI. They applied gradient boosted regression to regress multidimensional right ventricle shape landmarks from image appearance, which consider correlations between landmarks. Their method minimizes the shape alignment error over training data and shows better segmentation performance than multi-atlas-label-fusion based segmentation methods.

\begin{figure}
	\centering
	\includegraphics[width=3in]{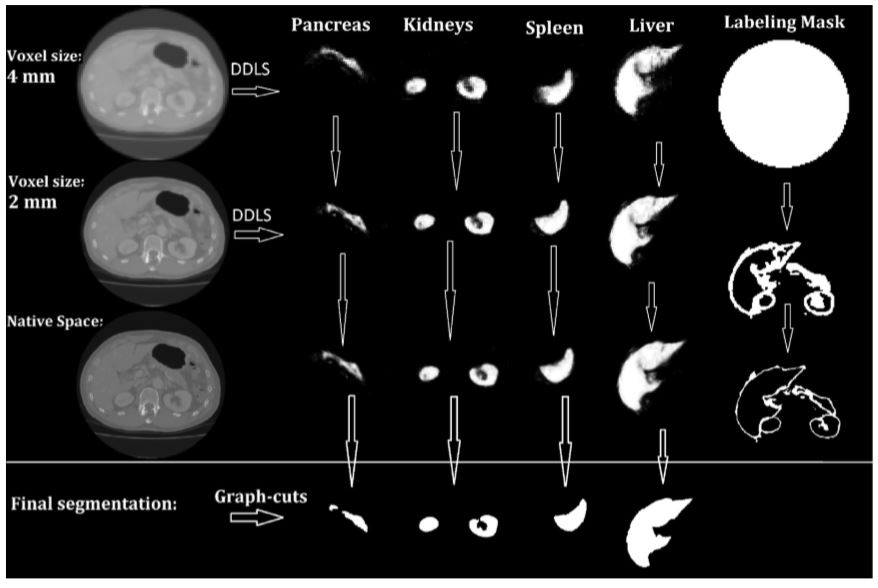}
	\caption{Tong et al. performed discriminative dictionary learning in muliresolution to generate probabilistic atlas for each organ. The graph-cuts algorithm is implemented in Native space, combining the information across resolutions and achieving the final segmentation results \protect\cite{Tong2015}. }
	\label{fig:tong}
\end{figure}

The other traditional supervised methods applied in segmentation tasks include dictionary learning and Bayes classification. Tong et al. proposed the extraction of voxel-intensity features for multi-organ segmentation (liver, kidneys, pancreas, and spleen) using dictionary learning and a sparse coding technique (Fig. \ref{fig:tong}) \cite{Tong2015}. The atlases selected against which to segment the images profoundly influence the performance of multi-based methods \cite{Aljabar2009}. To deal with the high inter-subject variation in CT images, they applied a voxel-wised local atlas selection strategy to improve performance. Griffis proposed a supervised learning method that automatically delineates stroke lesions using Na\"{\i}ve Bayes classification in single T1-weighted MRI sequence data \cite{Griffis2016}. In order to save time and money, their approach focuses on using single scan data, which detects direct lesion effects and has a better performance than manual delineation.

Image quality remains a limitation of the extraction of features from the radiology images. In many cases such as brain boundary segmentation, the data is of low contrast by nature. Additionally, both resolution and partial volume effects influence the definition of boundaries \cite{Wang2014a}. Some research contributions focus on multi-modalities to obtain complementary information \cite{Jin2014, ZhangDaoqiuang;Shen2013, Tong2014, Mitra2014}; however, it is difficult and inconvenient to apply various testing methods on patients.  Also, the accuracy of the segmentation system is difficult to measure and compare because the “ground truth” varies based on the delineation by different experts \cite{Eskildsen2012}. However, it is challenging and expensive to obtain manually labeled data from several experts for reliability tests\cite{Herrera}.

\begin{table}[]
	\centering
	\tiny
	\caption{Overview of segmentation methods for different radiological images  }
	\label{my-label}
	\begin{tabular}{llllll}
		\textbf{}                    & \textbf{image types} & \textbf{\# images} & \textbf{goal}      & \textbf{methods}                             & \textbf{Dice coefficients}                                                                              \\ \hline
		\cite{Roy2014}               & MRI                  & 12                 & Brain tissue       & Sparse dictionary learning                   & \begin{tabular}[c]{@{}l@{}}0.91 (Gray matter)\\ 0.87 (White matter)\end{tabular}                        \\
		\cite{Mitra2014}             & MRI                  & 36                 & Stroke lesion      & Random forest                                & 0.82                                                                                                    \\
		\cite{B2015}                 & CT                   & 42                 & Liver tumor        & Random forests \& supervoxels                & 0.93                                                                                                    \\
		\cite{li2015automatic}       & CT                   & 30                 & Liver tumor        & CNN                                          & 0.84                                                                                                    \\
		\cite{Liu2015}               & MRI                  & 70                 & Knee               & Multi-atlas context forests                  & \begin{tabular}[c]{@{}l@{}}0.97 (Bone)\\ 0.81 (Cartilage)\end{tabular}                                  \\
		\cite{Tong2015}              & CT                   & 150                & Multi-organ        & Discriminative dictionary learning           & \begin{tabular}[c]{@{}l@{}}0.90 (Liver)\\ 0.88 (Kidney)\\  0.55 (Pancreas)\\ 0.92 (Spleen)\end{tabular} \\
		\cite{Zhang2015}             & MRI                  & 10                 & Brain tissue       & CNN                                          & \begin{tabular}[c]{@{}l@{}}0.95 (Gray matter)\\ 0.86 (White matter)\end{tabular}                        \\
		\cite{roth2015deeporgan}     & CT                   & 82                 & Pancreas           & CNN                                          & 0.72                                                                                                    \\
		\cite{Griffis2016}           & MRI                  & 30                 & Stroke lesion      & Gaussian Na\"{\i}ve Bayes classification          & 0.81                                                                                                    \\
		\cite{ Si2016}               & MRI                  & 12                 & Brain lesion       & ANN                                          & 0.79                                                                                                    \\
		\cite{guo2016deformable}     & MRI                  & 66                 & Prostate           & Sparse auto-encoder \& sparse patch matching & 0.88                                                                                                    \\
		\cite{avendi2016combined}    & MRI                  & 45                 & Left ventricle     & CNN \& stacked-auto-encoder                  & 0.97                                                                                                    \\
		\cite{Kleesiek2016}          & MRI                  & 53                 & Brain tumor        & CNN                                          & 0.95                                                                                                    \\
		\cite{van2016combining}      & CT                   & 73                 & Lung texture       & Convolutional restricted Boltzmann machines  & 0.74                                                                                                    \\
		\cite{moeskops2016automatic} & MRI                  & 57                 & Brain segmentation & CNN                                          & 0.86                                                                                                    \\
		\cite{manniesing2017white}   & 4D-CT                & 22                 & Brain tissue       & SVM                                          & \begin{tabular}[c]{@{}l@{}}0.79 (Gray matter)\\ 0.81 (White matter)\end{tabular}                        \\
		\cite{havaei2017brain}       & MRI                  & 65                 & Brain lesion       & CNN                                          & 0.79                                                                                                    \\
		\cite{hu2016automatic}       & CT                   & 42                 & Liver tumor        & CNN                                          & 0.97                                                                                                    \\
		\cite{paredes2017deep}       & MRI                  & 73                 & Brain tumor        & CNN                                          & 0.65                                                                                                   
	\end{tabular}
\end{table}

\subsection{Computer Aided Diagnosis}
Computer-aided diagnosis (CAD) systems can detect, mark, and assess potential pathologies for radiologists to help improve identification accuracy in the case of data overload and human resource limitation. The analysis, quantification, and categorization of images with these methods is an important technique, which can improve patient safety and care. CAD systems have achieved breakthroughs in the detection of lesions \cite{OConnor2007, Yao2014}, epidural masses \cite{Liu2014}, fractures \cite{Yao2012}, as well as a degenerative disease \cite{Thung2014} and cancer \cite{Lehman2015}. Fisher's linear discriminant, Bayesian methods, artificial neural networks, and SVM are widely used as classifiers in CAD applications \cite{Yao2011, Dong2015}. Table \ref{tab:CAD} summarizes some current CAD investigations with machine learning techniques. 

\begin{table*}[]
	\centering
	\tiny
	\caption{A summary of recent CAD studies. \\AUC = area under curve; ROC = receiver operating cruve; TP = true positive rate; MAE = mean average error }
	\label{tab:CAD}
	\begin{tabular}{llllllll}
		\textbf{}                          & \multicolumn{1}{c}{\textbf{year}} & \multicolumn{1}{c}{\textbf{image type}} & \multicolumn{1}{c}{\textbf{\# cases}} & \multicolumn{1}{c}{\textbf{disease}} & \multicolumn{1}{c}{\textbf{Measurements}} & \multicolumn{1}{c}{\textbf{results}} & \multicolumn{1}{c}{\textbf{keywords}}                  \\ \hline
		\cite{Perez2014}                   & 2014                              & mammography                             & 956                                   & Breast cancer                        & AUC                                                   & 0.81                                 & Combination of classifiers                             \\
		\cite{Jiang2014}                   & 2014                              & mammography                             & 500                                   & Breast cancer                        & AUC                                                   & 0.91                                 & Na\"{\i}ve Bayes classification                             \\
		\cite{Torheim2014}                 & 2014                              & MRI                                     & 81                                    & Cervical cancer                      & Accuracy                                              & 0.69                                 & Texture features, SVM                                  \\
		\cite{Sun2015}                     & 2015                              & mammography                             & 340                                   & Breast cancer                        & AUC                                                   & 0.73                                 & Texture features, SVM                                  \\
		\cite{Perez2015}                   & 2015                              & mammography                             & 772                                   & Breast cancer                        & AUC                                                   & 0.89                                 & Feature selection method                               \\
		\cite{Wang2015a}                   & 2015                              & CT                                      & 750                                   & Lung                                 & AUC                                                   & 0.98                                 & Structured SVM                                         \\
		\cite{Antani2015}                  & 2015                              & X-ray                                   & 5,440                                 & Lung                                 & Accuracy                                              & 0.92                                 & SVM                                                    \\
		\cite{gopalakrishnan2015cmri}      & 2015                              & MRI                                     & 83                                    & Pediatric cardiomyopathy             & Accuracy                                              & 0.81                                 & Bayesian rule learning                                 \\
		\cite{Arevalo2015}                 & 2016                              & mammography                             & 736                                   & Breast cancer                        & AUC                                                   & 0.82                                 & CNN                                                    \\
		\cite{Singh2016}                   & 2016                              & mammography                             & 2,604                                 & Breast cancer                        & AUC                                                   & 0.93                                 & Wavelet neural network \\
		\cite{cheng2016computer}           & 2016                              & ultrasound                              & 520                                   & Breast lesions                       & Accuracy                                              & 0.82                                 & Stacked denoising auto-encoder                         \\
		\cite{rani2016detection}           & 2016                              & ultrasound                              & 95                                    & Liver lesions                        & Accuracy                                              & 0.87                                 & SVM                                                    \\
		\cite{Burns2016}                   & 2016                              & CT                                      & 104                                   & Vertebral body fractures             & TP                                                    & 0.81                                 & SVM                                                    \\
		\cite{ebsim2016detection}          & 2016                              & CT                                      & 409                                   & Wrist, radius, ulna fractures        & ROC                                                   & 0.89                                 & Random forest                                          \\
		\cite{Kooi2017}                    & 2017                              & mammography                             & 45,000                                & Breast cancer                        & AUC                                                   & 0.91                                 & CNN                                                    \\
		\cite{liu2017cade}                 & 2017                              & CT                                      & 1012                                  & Lung cancer                          & Sensitivity                                           & 0.89                                 & ANN                                                    \\
		\cite{miki2017classification}      & 2017                              & CT                                      & 52                                    & Teeth                                & Accuracy                                              & 0.89                                 & CNN                                                    \\
		\cite{mehrtasha2017classification} & 2017                              & CT                                      & 344                                   & Prostate cancer                      & ROC                                                   & 0.80                                 & CNN                                                    \\
		\cite{spampinato2017deep}          & 2017                              & X-ray                                   & 1391                                  & Bone age                             & MAE                                                   & 0.80                                 & CNN                                                    \\
		\cite{wang2017chestx}              & 2017                              & X-ray                                   & 108,948                               & Thorax diseases                    & Accuracy                                              & 0.63                                 & CNN                                                    \\
		\cite{rajpurkar2017chexnet}        & 2017                              & X-ray                                   & 112,120                               & Thorax diseases                   & AUC                                                   & 0.84                                 & CNN                                                    \\
		\cite{hsieh2017computer}        & 2017                              & MRI                                   & 107                              & Brain tumor                   & Accuracy                                                   & 0.88                                 & Logistic Regression                                                    		
	\end{tabular}
\end{table*}

Breast cancer is one of the most common cancers in the world. Currently, about one in ten women suffer from it, and early diagnosis and treatment of breast cancer could increase the chance of survival significantly \cite{Lee2010}. Among these diagnoses techniques, mammography is the best approach to detect breast cancer in its early stages and features indicating abnormalities can be extracted directly from medical images \cite{Nithya2011,Luo2012}. The identification of benign and malignant masses is the core principle for using mammography as a means to diagnose breast cancer \cite{Jiang1999}. Perez et al. developed machine learning classifiers that combine suitable feature selection methods with different machine learning techniques \cite{Perez2014}. The feature selection methods include chi-square discretization, information gain, one rule, relief, and u-test based filter. They then improved their feature selection algorithm called uFilter, which ranks features in a descending manner\cite{Perez2015}. Their method was useful for different datasets and reduced the number of employed features without decreasing the classification accuracies. 

The SVM classifier is widely used in breast cancer diagnosis with various features, such as wavelet features, gray-level-co-occurrence matrix features, intensity features, and texture features \cite{Sun2014, Sun2015}. Banaem et al. proposed a fully automatic tool that can classify the mammogram data into normal and abnormal. They used gray level co-occurrence and maximum difference method to extract proper features and the ensemble classification combining SVM, KNN and Na\"{\i}ve Bayes was applied to improve the diagnostic accuracy \cite{Banaem2015}. Many investigations not only consider the accuracy of the model, but also the model complexity. Arevalo et al. trained an SVM model that integrated two layers CNN for mass lesion classification \cite{Arevalo2015, Arevalo2015a}. Similarly, Jiao et al. trained two SVM classifiers using deep learning features extracted from two different layers of CNN networks \cite{Jiao2016}. An automated CAD system was proposed, combing the content-based image retrieval to detect masses \cite{Jiang2014}. The main idea of their approach is to use scale invariant feature transform features to match query mammogram and exemplar masses in the database, and then uses Na\"{\i}ve Bayes classification and thresholded maps to detect masses. In their method, the model complexity is low as there is no sliding window-based scanning.

The SVM method is also widely studied in other diagnose such as lesion, injury and fractures detection. In these diagnosis tasks, choice of features plays a significant role in model accuracy. Torheim et al. predicted cervical cancer from dynamic contrast enhanced MRI. In their study, gray-level-co-occurrence matrices based textural features were implemented as explanatory variables \cite{Torheim2014}. Wang improved the accuracy of lung lesion detection from CT images by using a 3D matrix patterns-based SVM with latent variables. Their study focused on detecting lung lesions that had irregular shape and low-intensity, rather than the nodules, which provides a new thought for the detection of lung lesions \cite{Wang2015a}. In the detection task of thoracic and lumbar vertebral fractures \cite{Burns2016}, Burns extracted 28 features from the cortical shell from CT images based on the essential method (Denis ‘middle column’), which is specific to detection of fracture discontinuities on vertebral body cortices. Jin et al. established a prognosis model of cervical spondylotic myelopathy using a least-square SVM \cite{jin2016machine}. In their studies, they extracted values of fractional anisotropy, axial diffusivity, mean diffusivity and radial diffusivity from each slice of DTI metrics as features, which yielded 88.62\% prediction accuracy.  

The popular methods such as deep relief networks \cite{abdel2016breast} and convolutional neural networks \cite{wang2016deep,cheng2016computer} achieved promising results in many diagnosis applications. The important diagnosis tasks based on neural networks include chest pathology identification \cite{ bar2015deep}, cancer detection\cite{rasti2017breast,wang2017searching} and lung diseases \cite{ anthimopoulos2016lung} . Neural network based methods rely heavily on the support of big data. A semi-supervised algorithm has been proposed to deal with a large amount of unlabeled data with CNN approaches \cite{Sun2016}. Their approaches using unlabeled data increased the overall accuracy, rather than just using labeled data.

There are many advantages to using machine learning techniques in CAD systems. The first advantage of machine learning is its accurate and robust performance in many radiology studies. For instance, CAD systems have reached perfect accuracy e.g., over 99$\%$ in oral cancer detection\cite{Exarchos2012}, which is comparable to manual diagnosis. Moreover, CAD systems are expected to perform consistently and produce robust results with large amounts of data at any time and in any space, while manual diagnosis results may be affected by fatigue, reading time, and emotion on the part of the practitioner. The second advantage is that the diagnosis can be finalized in a brief time. Many radiology analyses are time-consuming and require experienced radiologists. For example, the software developed for breast cancer prediction\cite{Patel2016} can review charts 30 times faster than humans can. Another example is that the suggested approach in breast cancer diagnosis is the double reading of mammograms by two radiologists \cite{Perez2014}. With the help of a CAD system, only one radiologist is needed instead of two, which could help to increase the survival rate among women in a cost-effective manner \cite{Ayer2010a}. 

 Although we are witnessing better accuracy of computer-aided diagnosis systems to tackle the most common clinical problems, current contributions still have potentials for improvement before their applications in clinical practice. First, a majority of current diagnosis contributions mainly focus on the prediction of one type of disease, which may not meet the clinical demands. There may be one or more diseases existing in one radiological image (for example, effusion \& atelectasis in one chest x-ray image). Second, the current model trainings is mainly based on one type of measurement.  However, most disease decisions in clinical practice rely on multiple domain measurement (such as patient demographics, image screening, blood test and drug test).  Information from multi measurement may increase model accuracies. Third, current medical datasets mainly cover common diseases. Only a limited number of rare diseases are exposed to human clinicians, and many contributions may not consider these individual cases during their model training. More comprehensive systems that can detect various types of diseases and report rare cases are expected to be seen in the future.
\subsection{Functional Brain Studies and Neurological Diseases}
Brain tumors, neurological disorders such as epilepsy, and neurodegenerative diseases have attracted much attention in brain-related investigations. In brain-related image diagnosis, a large number of features can be extracted from brain regions related to the nature of pathological changes. Cortical thickness \cite{Kloppel2008}, the volume of brain structures \cite{Chupin2009}, and voxel tissue probability maps around some regions of interest \cite{Fan2007} are popular choices for feature extraction \cite{Chen2015}. Different MRI modalities such as T1-weighted or fluid-attenuated inversion recovery imaging contain large amounts of information and noise \cite{Si2016}. Therefore, compelling feature fusion strategy is necessary for neuroimaging analysis and classification \cite{Liu2014a, Chu2012}.
\subsubsection{Support vector machine in brain studies}
\begin{figure}
	\centering
	\includegraphics[width=2.5in]{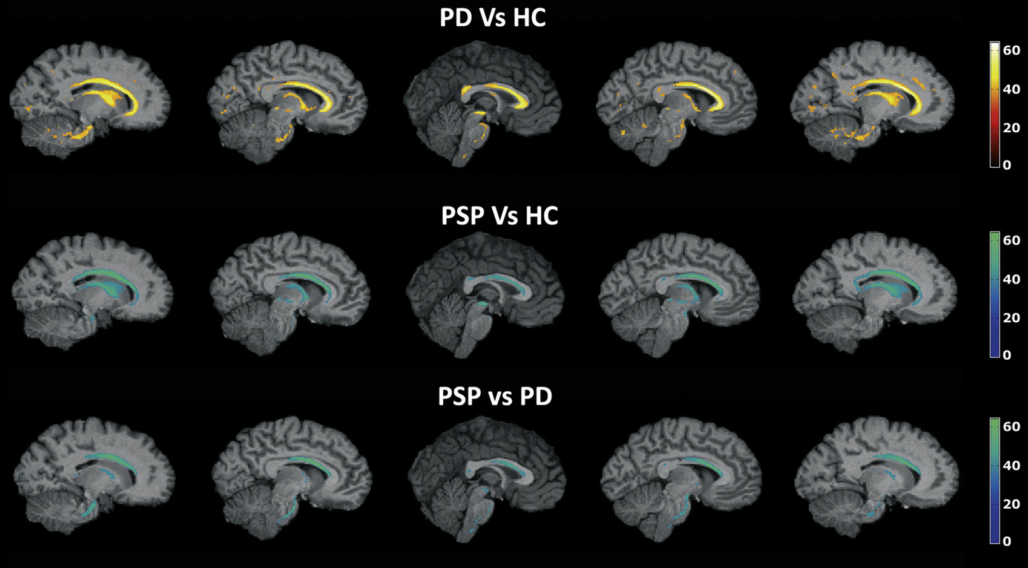}
	\caption{Salvatore et al. \protect\cite{Salvatore2014} proposed a supervised learning method to identify PD and PSP using MR images. The figures show maps of voxel-based pattern distribution of brain structural differences. The color scale expresses the importance of each voxel in SVM classification.}
	\label{fig:Salvatore}
\end{figure}

In brain studies, the SVM is a powerful tool for feature selection, which may improve model accuracy. Larroza et al. developed a classification model of brain metastasis and radiation necrosis in contrast-enhanced T1-weighted images. Features were extracted by texture analysis and reduced by using a linear SVM \cite{Larroza2015}.  Bron proposed a feature selection method based on the SVM significance value \cite{Bron2014}. The significance value (p-value) serves to quantify the contribution of each feature to the SVM classifier and is used to reduce voxel-based morphometry features.
Neurodegenerative diseases such as Parkinson's disease begins before the onset of symptoms. Thus, medical treatment is more effective if it is detected in early stage.  Among various forms of Parkinsonism, progressive supranuclear palsy is one of the most difficult to be identified in an early disease stages \cite{Gelb1999}. Salvatore et al. proposed to classify control subjects, progressive supranuclear palsy patients, and Parkinson's disease patients based on SVM models. Features were extracted by spatial transformations and principal component analysis from T1-weighted sequences. The accuracy of discrimination of Parkinson's disease and progressive supranuclear palsy is above 90$\%$\cite{Salvatore2014}. Fig. \ref{fig:Salvatore} uses a color scale to express the importance of each region during classification. To improve the diagnostic accuracy of classifying Parkinson's disease patients, Singh proposed an unsupervised feature extraction method from a T1-weighted sequence by using a Kohonen self-organizing map algorithm. With the least square SVM, the accuracy of identifying the affected area in Parkinson's disease is up to 99$\%$ \cite{Singh2015}. In \cite{Chen2015}, features were extracted using a deep network and a stacked denoising sparse autoencoder, which makes the input data points more linearly separable in SVM \cite{Chen2015}.  Liu proposed an inherent structure-guided multi-view learning method to classify Alzheimer's disease and mild cognitive impairment patients \cite{Liu2015a}. They extracted 1500 features from gray matter density, and multi-task feature selection was applied to reduce the dimension, followed by an ensemble classification method using multiple SVM classifiers. 

Besides the disease studies, some research work applied machine learning techniques to understand the brain's functional network architecture. Smyser compared the fMRI data from 50 preterm-born and 50 term-born infants using SVM \cite{Smyser2016}. Their results show that inter- and intra-hemispheric functional connections throughout the brain are stronger in full-term infants. Their findings might be helpful for the development of models for defining indices of brain maturation.
\subsubsection{Ensemble learning in brain studies}
\begin{figure}
	\centering
	\includegraphics[width=4in]{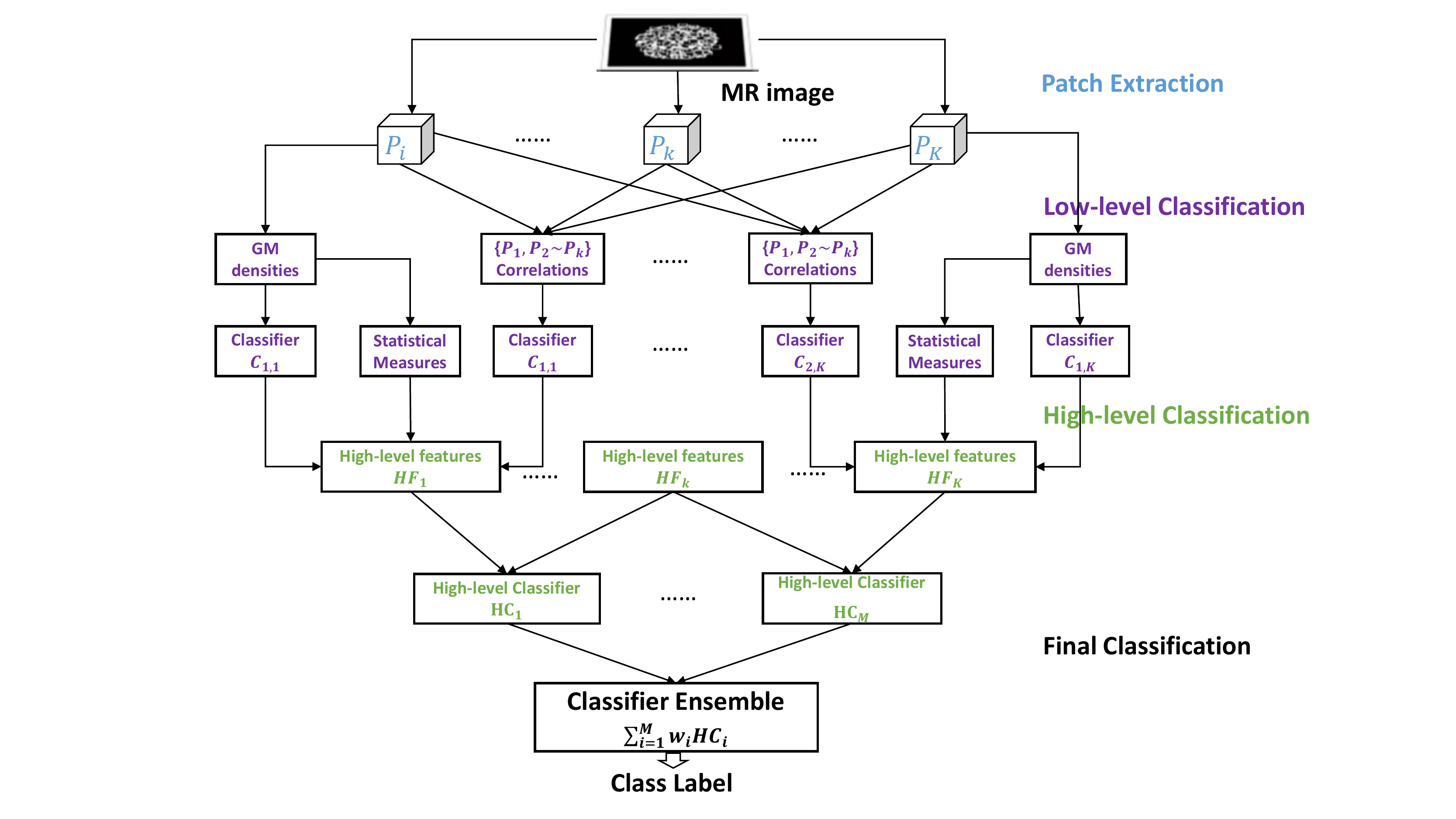}
	\caption{Flow chart of the hierarchical classification algorithm proposed in \protect\cite{Liu2014a}, the low-level classifiers are used to transform imaging and spatial-correlation features from the local patch, and the output of these low-level classifiers is integrated into high-level classifiers with coarse-scale imaging features. The final classification is achieved by ensemble outputs from high-level classifiers. }
	\label{fig:Liu}
\end{figure}

 Ensemble learning methods combines multiple classifiers, which is popular in Alzheimer's disease diagnosis. Alzheimer's disease is estimated to affect around 5.4 million patients in America, and is the most common form of dementia among the elderly population \cite{AlzheimersAssociation2015, Thung2014}, which leads to the loss of cognitive function and death. Liu proposed a classification framework that works on different image modalities for the classification of Alzheimer's disease patients \cite{Liu2014a}. Their method contains level classifiers: low-level classifiers that use different types of low-level features from patches, high-level classifiers that combine coarse-scale imaging features in each patch and outputs of low-level classifiers, as well as a final ensemble classification that combines the decisions of a high-level classifier with a weighted voting strategy (Fig. \ref{fig:Liu}).  In \cite{Li2014}, high accuracy results were obtained from Alzheimer's disease/healthy and mild cognitive impairment/healthy classification. However, accuracies in classifying mild cognitive impairment as converted to Alzheimer's disease are very low (57.4$\%$), but is slightly higher than majority classification. Komlagan et al. developed an ensemble learning method using gray matter for a weak classifier and selecting the most relevant sub-ensembles through sparse logistic regression \cite{Komlagan2014}. They trained a global linear SVM classifier for the final classification. Combining high quality biomarkers with advanced learning methods makes results comparable to those of multi-modality methods. 

\subsubsection{Others techniques in brain studies}
Some researchers leverage regression and principle analysis components in classification and feature mapping.  Ahmed et al. detected neocortical structural lesions with an automated approach, which contained five surface-based MRI features and combined them in a logistic regression \cite{Ahmed2015}. To deal with imbalance issues, they used a ``bagging'' approach and an iterative-reweighted least squares algorithm. The base-level classifier was trained on all the minority class instances and the same size of random data from majority class instances.  Hong proposed a machine learning technique combining surface-based analysis in patients with a subtype of focal cortical dysplasia \cite{Hong2014}. Their automated approach used features of Focal cortical dysplasia morphology and intensities, and Fisher's linear discriminant was applied as a classifier to identify Focal cortical dysplasia in patients.   
Huang proposed the use of a soft-split random forest to predict clinical scores in Alzheimer's disease patients \cite{B2015a}. In their method, lasso regression is applied to map MRI features, and then features are reduced by principal component analysis. Li combines principal component analysis, the lasso method, and a deep learning framework to extract features by fusing information from MRI and PET images in the classifier of Alzheimer's disease/mild cognitive impairment patients \cite{Li2014}.
Zhu et al. focused on the identification of Alzheimer's disease patients with multi-view or visual features of image data. They proposed several feature selection approaches for Alzheimer's disease classification. They integrated subspace learning into a sparse least square regression framework for multi-classification in 2014 \cite{Zhu2014}. Then, they mapped the histogram of oriented gradient features (which are diverse) onto a region of interest features (which are robust to noise), which provided complementary information for features and enhanced disease status identification accuracy \cite{Zhu2015}. Other machine learning techniques such as convolutional neural networks are widely investigated in the field as well. Table \ref{tab:brain} summarizes recent contributions related to Alzheimer's disease classification.

\begin{table}[]
	\centering
	\tiny
	\caption{Recent studies on Alzheimer's diseases \\ NC: normal; AD: Alzheimer's disease; pMCI: progressive mild cognitive impairment; sMCI: stable mild cognitive impairment}
	\label{tab:brain}
	\begin{tabular}{llllllll}
		\textbf{}                            & \textbf{year}         & \textbf{databses}               & \textbf{image \#}     & \textbf{image types}     & \textbf{classification gruops} & accuracy & keywords                                        \\ \hline
		\multirow{2}{*}{\cite{Tong2014}}     & \multirow{2}{*}{2014} & \multirow{2}{*}{ADNI}           & \multirow{2}{*}{834}  & \multirow{2}{*}{MRI}     & AD vs. NC                      & 89\%     & \multirow{2}{*}{Multiple instance learning}     \\
		&                       &                                 &                       &                          & pMCI vs. sMCI                  & 70\%     &                                                 \\
		\multirow{2}{*}{\cite{Zhu2014}}      & \multirow{2}{*}{2014} & \multirow{2}{*}{ADNI}           & \multirow{2}{*}{202}  & \multirow{2}{*}{MRI+PET} & AD vs. MCI vs. NC              & 73.35\%  & Sparse discrimination feature selection         \\
		&                       &                                 &                       &                          & AD vs. pMCI vs. sMCI vs. NC    & 61.06\%  &                                                 \\
		\multirow{2}{*}{\cite{Guerrero2014}} & \multirow{2}{*}{2014} & \multirow{2}{*}{ADNI}           & \multirow{2}{*}{1071} & \multirow{2}{*}{MRI}     & AD vs. NC                      & 89\%     & \multirow{2}{*}{Manifold and transfer learning} \\
		&                       &                                 &                       &                          & pMCI vs. sMCI                  & 73\%     &                                                 \\
		\cite{Komlagan2014}                  & 2014                  & ADNI                            & 814                   & MRI                      & pMCI vs. sMCI                  & 75.6\%   & Gray matter grading, weak-classifier fusion     \\
		\multirow{3}{*}{\cite{Liu2015a}}     & \multirow{3}{*}{2015} & \multirow{3}{*}{ADNI}           & \multirow{3}{*}{459}  & \multirow{3}{*}{MRI}     & AD vs. NC                      & 93.83\%  & \multirow{3}{*}{Hierarchial fusion of features} \\
		&                       &                                 &                       &                          & pMCI vs. sMCI                  & 80.9\%   &                                                 \\
		&                       &                                 &                       &                          & pMCI vs. NC                    & 89.09\%  &                                                 \\
		\cite {Cheng2015}                    & 2015                  & ADNI                            & 202                   & PET+ MRI                 & pMCI vs. sMCI                  & 78.7\%   & Multimodel multi-label transfer learning        \\
		\multirow{3}{*}{\cite{Zhu2015}}      & \multirow{3}{*}{2015} & \multirow{3}{*}{ADNI}           & \multirow{3}{*}{830}  & \multirow{3}{*}{MRI}     & AD vs. NC                      & 91.31\%  & \multirow{3}{*}{HoG mapping}                    \\
		&                       &                                 &                       &                          & MCI vs. NC                     & 78.07\%  &                                                 \\
		&                       &                                 &                       &                          & pMCI vs. sMCI                  & 75.54\%  &                                                 \\
		\cite{Keserwani2016}                 & 2016                  & OASIS                           & 416                   & MRI                      & AD vs. NC                      & 80.76\%  & Gabor filter                                    \\
		\cite{sarraf2016deepad}              & 2016                  & ADNI                            & 416                   & MRI                      & AD vs. NC  vs. MCI             & 89.1 \%  & CNN                                             \\
		\cite{Long2016}                      & 2016                  & Self-collected                  & 67                    & MRI                      & AD vs. NC                      & 96.77\%  & SVM                                             \\
		\cite{Komlagan2014}                  & 2014                  & ADNI                            & 814                   & MRI                      & pMCI vs. sMCI                  & 75.6\%   & Gray matter grading, weak-classifier fusion     \\
		\multirow{3}{*}{\cite{khazaee2016application}}             & \multirow{3}{*}{2016} & \multirow{3}{*}{Self-collected} & \multirow{3}{*}{89}   & \multirow{3}{*}{fMRI}    & AD vs. NC                      & 97.50\%  & \multirow{3}{*}{SVM}                            \\
		&                       &                                 &                       &                          & MCI vs. AD                     & 87.30\%  &                                                 \\
		&                       &                                 &                       &                          & MCI vs. NC                     & 72.00\%  &                                                 \\
		\cite{Armananzas2016}                & 2016                  & Dartmouth College               & 116                   & MRI                      & AD vs. NC                      & 97.14\%  & Feature ranking selection                       \\
		\cite{sarraf2016deep}                & 2016                  & Self-collected               & 43                   & fMRI                     & AD vs. NC                      & 96.85\%  & Deep learning selection                       \\
		\cite{schouten2017individual}                & 2017                  & Self-collected               & 250                   & DTI                      & AD vs. NC                      & 89.60\%  & Elastic net selection                       \\
		&                       &                                 &                       &                          &                                &          &                                                
	\end{tabular}
\end{table}

\subsection{Image Retrieval}
\begin{figure}
	\centering
	\includegraphics[width=3in]{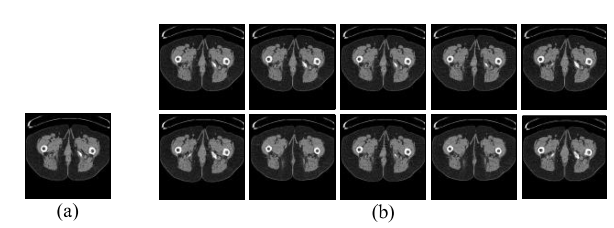}
	\caption{The method for retrieving images using Local wavelet pattern features and similarity measurement. All retrieved images are from the same category, achieving 100 \% precision in this example \protect\cite{Dubey2015}: (a) Query image. (b) Top 10 retrieved images. }
	\label{fig:Dubey}
\end{figure}

With the increased use of modern medical diagnostic techniques, there are numbers of medical images stored in hospital archives. Manual annotation and attribution of these images are impractical \cite{Ibrahim2016}. Picture archiving and communication systems have been widely introduced in many hospitals \cite{Kachhela2013}. These systems could retrieve images based on keywords, however these images may not be directly useful in helping to making clinical decisions. Different from traditional image search systems, which are based on matching keywords and image tags, content-based image retrieval extracts rich contents from images and searches for other images with similar contents. Content-based image retrieval is becoming necessary for the medical image databases, which may potentially become effective tools of anatomical and functional information for diagnostic, educational, and research purposes \cite{Wei2006}. Table \ref{tab:image-retrieval} lists current investigations on image retrieval.  
Recently, similarity or distance learning is a hot topic in the image retrieval field. Traditional choices include the Euclidean distance function, $x^2$ square distance function, Mahalanobis distance, $l_1$ norm distance function \cite{Meng2016}, maximum likelihood approach \cite{Yu2008} and Bayes ensemble \cite{Emrich2010}. Like other machine learning tasks, features extraction is an important step in image retrieval systems. Kurtz et al. proposed the use of hierarchical semantic-based distance to retrieve images based on 72 manually annotated semantic terms from each region of interest \cite{Kurtz2014}. Then, they built a semantic framework that learns image descriptions of each term using Riesz wavelets and SVM.  In \cite{Dubey2015}, local wavelet patterns were introduced as  a new feature descriptor. Their experiments first utilized the relationship among the neighboring pixels and performed well in CT image retrieval. Their results were shown in Fig. \ref{fig:Dubey}. Different from traditional similarity learning that only maximizes the margin, Meng et al. proposed a novel similarity learning algorithm which considered the top precision performance measure in the loss function \cite{Meng2016}. Their methods showed advantages over other traditional similarity learning methods.

The other supervised techniques applied in retrieval systems include online dictionary learning, ensemble learning and principal component analysis. The main advantage of an online dictionary learning system is the computational time, as learned dictionaries are used to represent the dataset in a sparse model, which is an valuable tool for representing data \cite{Ramirez2010}. A method using online dictionary learning and its features extracted by multi-scale wavelet packet decomposition from different types of images is proposed in \cite{Srinivas2014}. Srinivas et al. proposed a medical image classification approach using online dictionary learning with the edge- and patch- based features to distinguish 18 categories \cite{Srinivas2016}.  Ahn et al. developed a robust method to improve X-ray image classification \cite{Ahn2016}. A fusion strategy was proposed that combines domain transferred convolutional neural networks and sparse spatial pyramid classification. The combined method performs better than the single method used. Faria et al. proposed a retrieval method for brain MRI images. They captured anatomical features from T1-weighted images using least-square discriminant analysis and principal component analysis and performed a search for images between healthy controls and patients with primary progressive aphasia \cite{Faria2015}. 

As large portions of medical images in the dataset lack labels and annotations, semi-supervised and unsupervised techniques are required in the retrieval systems. Uunsupervised image retrieval based on the clustering method using K-SVD executes iterations between grouping similar images into clusters and generating a dictionary for clusters until clusters converge \cite{Srinivas2015}. The advantage of this method is that it requires no training data for classification and is not restricted to a specific context. Since labeled data is limited, Herrera proposed a semi-supervised learning method for image classification using k-nearest neighbors to expand the training data set and a random forest for final classification \cite{Herrera}. 

Medical image retrieval gives an opportunity for clinicians to search for similar disease cases. Accuracy and performance time are both vital aspects of the medical image retrieval system. A practical model and relevant image feature extraction are required to get better results. Furthermore, some image retrieval contributions mainly investigated small datasets and limited disease cases. With an increasing number of digital radiological images in hospital databases every year, whether systems can retrieve disease cases stably and efficiently in huge datasets is still an exciting avenue for researchers.

\begin{table*}[]
	\centering
	\tiny
	\caption{A summary of recent image retrieval research using machine learning techniques}
	\label{tab:image-retrieval}
	\begin{tabular}{llllll}
		& year & image types   & \# images                                                                                                     & results                                                            & keywords                                          \\ \hline
		\cite{Kurtz2014a}   & 2014 & CT            & 72                                                                                                           & AUC:0.93                                                           & Riesz wavelets, hierarchical semantic-based distance                     \\
		\cite{Faria2015}    & 2015 & MRI           & 30                                                                                                           & Accuracy:0.88                                                     & Partial least square discriminant analysis, principal component analysis   \\
		\cite{Dubey2015}    & 2015 & CT            & \begin{tabular}[c]{@{}l@{}}EXACT09:40\\ TCIA: 604\end{tabular}                                                 & \begin{tabular}[c]{@{}l@{}}Precision:\\ 0.88\end{tabular} & Local wavelet pattern                             \\
		\cite{Cao2015}      & 2015 & Multimodality & \begin{tabular}[c]{@{}l@{}}ImageCLEF:\\ 10 thousand\end{tabular}                                             & MAP:0.29                                                          & Deep Boltzmann machine                            \\
		\cite{Verma2015}    & 2015 & MRI           & OASIS:421                                                                                                    & Precision: 0.48                                                  & Local binary patterns, gray-level-co-occurrence matrices                                          \\
		\cite{Srinivas2016} & 2016 & X-ray \& CT  & ImageCLEF:5400                                                                                               & Accuracy:0.98                                                     & Sparse representation, online dictionary learning \\
		\cite{Meng2016}     & 2016 & Multimodality & \begin{tabular}[c]{@{}l@{}}Indoor:15620,\\ Caltech256:30670,\\ Corel5000:5000,\\ ImageCLEF:2785\end{tabular} & \begin{tabular}[c]{@{}l@{}}Top precision:\\ 0.36\end{tabular}      & Support top irrelevant machine                  \\
		\cite{lan2018simple}    & 2017 & CT            & \begin{tabular}[c]{@{}l@{}}EXACT09:675\\ TCIA: 604\end{tabular}                                                 & \begin{tabular}[c]{@{}l@{}}Precision:\\ 0.96\end{tabular} & Gabor and Schmid filters                                           
	\end{tabular}
\end{table*}

\subsection{Image Prediction}
\begin{figure}
	\centering
	\includegraphics[width=3in]{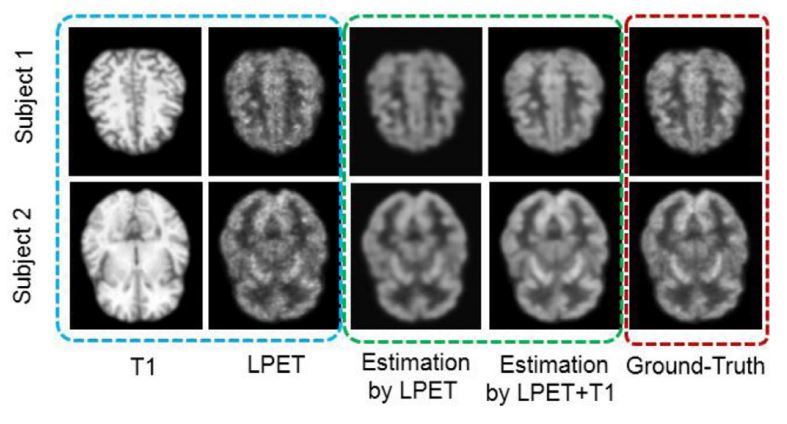}
	\caption{Deep auto-context convolutional neural networks were proposed for standard-dose PET (SPET) image estimation from low-dose PET (LPET) and T1 images \protect\cite{xiang2017deep}.
SPET images were estimated by using LPET images along and combination of LPET and T1 images. The neural networks perform better results when two image modalities were included.}
	\label{fig:xiang}
\end{figure}

With the development of neuroimaging techniques, various new image modalities have been applied in daily clinical practice to make diagnosis and treatment more efficient and accurate.  Thus, image prediction methods, which combine various image modalities and provide information for diagnosis, are fundamental. The main idea of image prediction is to estimate radiological images in different modalities or higher resolution, which can provide detailed functional information for assessment and diagnosis. 
PET is a molecular imaging technique which is widely used in clinical cancer diagnosis that produces 3D images, which can reflect tissue metabolic activity in the human body \cite{Rohren2004}. However, the quality of PET images is proportional to the dose injected and imaging time. As a result, low-dose PET images can suffer in quality. Thus, a great deal of effort has been made to predict high-quality PET images. Kang proposed a regression forest based approach to predict standard-dose PET images from low-dose PET and multimodal MRI images \cite{Kang2015}. They used a regression forest as their non-linear prediction model and features from local intensity patches of MRI data and low-dose PET. Meanwhile, Wang used a mapping-based sparse representation approach for prediction \cite{Wang2016a}. They used a graph-based distribution mapping method to reduce the patch distribution differences between MRI and low-dose PET and constructed a patch selection based dictionary learning method to predict standard-dose PET. Both methods performed better when compared with a path-based sparse model. In \cite{xiang2017deep}, Xiang et al. used convolutional neural networks to estimate standard PET image from low dose PET/MR image (Fig. \ref{fig:xiang}). By using neural network techniques, they can map the inputs to the output directly, without any pre/post-processing beyond the optimization in the training stage. Huynh predicted CT images from MRI data using a structured random forest instead of a classical random forest \cite{TriHuynh2015}. A structured random forest is an extension of a random forest, which predicts structured outputs instead of scalar outputs \cite{Kontschieder2011, Dollar2013}. Characterizing the information obtained from multiple sources of information improves prediction accuracy.

Compared to classification and segmentation tasks, contributions involving radiological imaging prediction is still quite limited. The rapid developments in hybrid imaging scanners (PET-CT, SPECT-CT, PET-MRI) has provided integrated images for diagnostic purposes. The primary challenge remains to find the appropriate way to match the correspondences among different images modalities \cite{yang2017quicksilver}. Besides, current contributions are mainly focused on brain data, but we expect to see more non brain related contributions to this area soon.

\section{Current Challenges} 
Until now, machine learning has been used to help radiologists in diagnostic tasks, but still cannot be a substitute for the clinician's role. There are some limitations regarding the application of machine learning in clinical practice \cite{Cerasa2015}. One such limitation is that the majority of these studies in radiology are based on supervised learning. The algorithms learn specific patterns based on previous decisions made by radiologists. The algorithm is expected to reach an accuracy of 100$\%$ compared with the human clinician, however, in many cases, an accurate diagnosis was made by several radiologists after multiple diagnostic tests. Whether a machine can perform well alone still needs to be investigated in the future. 

Facilitating data collection and sharing is a crucial point for further investigation in many studies. Clinically applied algorithms depend on two critical factors: robustness for large datasets and accuracy achieved \cite{Weese2016}. In many machine learning related studies, the size of the chosen dataset is relatively small \cite{Zhang2015, Faria2015, Griffis2016, Maier2015, Chen2014a}. This is due to limited access to patients or limited diagnostic work in the research setting. For example, in some clinical practices, there is no pathological exam done as a follow-up procedure in cases of brain metastasis \cite{Larroza2015}, which limits the acquisition of data. The application of machine learning to a limited number of cases (only around twenty patients were studied in some cases) is not always persuasive. Whether these clinical tools are robust enough to analyze immense amounts of medical data accurately remains a question. While 20-50 images were sufficient in past research, hundreds of or more image sets are required in the future to meet increasing requirements concerning robustness and accuracy. The creation of large databases and sharing centers such as ADNI for Alzheimer's disease patients, NIH repositories for chest X-rays, and TCIA for cancer imaging help to effectively collect millions of images for research. Furthermore, current studies trained and evaluated their models based on various datasets, which makes it challenging to compare their algorithms. A systematic evaluation standard based on various diseases and various public datasets is required in medical applications.

Excepting the data size, data quality and feature selection remain highly important for effective machine learning techniques. Medical images contain rich features that are clinically important. It is challenging to use low-level image features to get the visual appearance of disease. However, high dimensional features could be redundant for results in many fields such as image retrieval and classification. The choice of different high-level descriptions as input features is a prominent research topic. Choosing informative features for training can lead to robust models, whereas overfitting, underfitting, and misclassification usually occur when features are not selected well. More work remains to be done for selecting and utilizing proper features from images.

Transfer learning is an accepted scenario to learn information from limited datasets. There are three main transfer learning research directions in the field of radiological imaging. One is used to reduce bias among different equipment for image acquisition and different protocols \cite{paredes2017deep,cheplygina2016asymmetric}. The second approach is to learn various abnormalities from the same data source \cite{shen2016learning,cheng2015multimodal}. The last approach is to find a good feature representation from various domains and then apply them to the radiological imaging field \cite{paul2016deep}. Transfer learning allows us to deal with various scenarios by leveraging the already existing information from some related task or domain. For more details on transfer learning techniques and radiological imaging, readers should refer to \cite{cheplygina2018not}.

Imbalanced data is standard in medical diagnosis, where the majority of data is normal data, but only a minority class is abnormal \cite{Mena2006a}. For example, brain tumors are not common, occurring only in 1 $\permil$ of the population. However it remains the most fatal form of cancer \cite{Bauer2013}. This data imbalance might affect prediction accuracy and cause a bias toward the majority class \cite{Japkowicz2002}. Several researchers considered the imbalanced situation in their models \cite{Ahmed2015,Meng2016,Zhang2015}, however, the majority of studies have not properly addressed this issue \cite{Zhang2015, Wang2016}, or they use the same amount of data from different classes. How to utilize imbalanced data to improve the accuracy of machine learning algorithms remains an open question in this field. 

Large amounts of radiological images are produced in hospitals every year. However, most of these images are not utilized for further training of machine learning algorithms, as the training process constrains available resources. Useful information is hidden in this mass of data, and diagnostic machine learning models could be improved by using these streaming data. Online learning is a novel idea in recommender systems and other machine learning based systems that could update the model while streaming data are currently developed in other fields. This idea can also be transferred to medical diagnosis systems to make full use of streaming image datasets.   

Researchers have generated more powerful and more innovative diagnosis models for radiological imaging \cite{samek2017explainable}. However, very few of them are commercialized and deployed into the market. 
The main challenge is to comply with government requirements in various countries \cite{thelisson2017regulatory}.  Current FDA protocols suggest that medical products should pass the clinical trials, and be produced, commercialized, and used in a defined, unchanging form. If a machine learning model is used for medical diagnosis, a pre-build and freeze model must be tested in different clinical environments, assessed for various real-life medical conditions, and carefully evaluated on how these conditions affect the accuracy of the diagnosis. A recent study showed that a pre-trained model demonstrated significantly lower external performance on the data obtained from another hospital system \cite{zech2018variable}. In order to reduce the bias and improve performance, current machine learning solutions in nonmedical fields typically update the parameters of the models every time new data is included. However, this is not realistic for medical products as the system must pass a new clinical trial after the update. Therefore, this remains a major issue for machine learning algorithms in medical applications.

Knowing how deep neural networks work is an open question, and this prompts clinicians and patients to distrust these models. Due to huge amounts of parameters in the models, it is difficult to interpret how the models make diagnostic decisions between input and output. This could potentially be fatal if a machine learning model leads to a wrong conclusion \cite{caruana2015intelligible}, as medical experts can not verify these models. Deep learning researchers have recently computed heat maps using class activation maps in order to a more concrete analysis of how these models perform \cite{zhou2016learning}. An activation map visually highlights the discriminative regions in medical images that models used to identify the category. However, the related researches on network explainability and visualization are still limited in the field of radiological imaging.

 \section{Conclusion and Future Work\label{sec: conclusion}}
 In this paper, we reviewed five applications of machine learning techniques on radiologic images: image segmentation, computer-aided detection and diagnosis, functional brain studies and neurological disease diagnosis, image classification and retrieval, ands image registration. While machine learning techniques are active in computer-aided systems to assist radiologists in daily diagnosis and studies, the use of machine learning techniques in radiology is still evolving. There are many strategies that this field could investigate in the future：
 \begin{itemize}
 \item[-]Previous contributions have shown that machine learning-based systems showed accurate results comparable to those of radiologists themselves. However, the system accuracy of these techniques must still be improved, that is, systems must be more accurate than those of radiologists. Otherwise, the widespread application of machine learning techniques will be limited. A possible approach to achieve this superior performance is to design  better machine learning models or to gather more representative data that can be continuously used to improve the algorithms.
 
 \item[-]Although the core advancement of deep learning is its ability to learn useful features directly from data, its accuracy and performance are highly limited by the size of data. Traditional machine learning methods still play a role in the case of small amounts of labeled data. Understanding how to choose and use features from images effectively is still a significant direction for these traditional methods. 
 
 \item[-]Another issue deals with the translation of these techniques to clinical practice. While many machine learning algorithms have already shown good results, it still needs to pass clinical trials required by the government. Additionally, many people still believe in human decisions, as clinicians always consciously tend to decide with all the relevant information in mind. These decisions make it difficult to justify the use of algorithms for clinical decision making in all possible cases, but through rigorous research contributions, we can justify the use of machine learning algorithms in the cases when patients’ outcomes can be improved.
\end{itemize}
 The application and development of machine learning techniques to radiological images is a hot topic currently and a large number of algorithms are being developed to ensure higher accuracy and lower computational complexity. We expect that machine learning techniques will become essential components in clinical tools and will be widely used to assess patients' health in the future.  
 
\section*{Acknowledgment}
Research reported in this publication was supported by the Eunice Kennedy Shriver National Institute of Child Health \& Human Development of the National Institutes of Health under Award Number R01HD092239. The content is solely the responsibility of the authors and does not necessarily represent the official views of the National Institutes of Health.

\section*{Conflicts of Interest}
None declared.
\bibliographystyle{plain}

\end{document}